\documentclass{article}
\usepackage[utf8]{inputenc}
\usepackage[english]{babel}
\usepackage{graphicx}
\usepackage{hyperref} 
\usepackage{geometry}
\usepackage{amsmath}
\usepackage{amssymb}
\usepackage{amsfonts}
\usepackage{multicol}
\usepackage{float}
\usepackage{caption}
\usepackage{subcaption}
\usepackage{cite}
\usepackage{amsthm}
\usepackage[utf8]{inputenc}
\usepackage{xcolor}
\def\@fmsl@sh#1#2#3{\m@th\ooalign{$\hfil#1\mkern#2/\hfil$\crcr$#1#3$}}
 \def\eq#1\en{\begin{equation}#1\end{equation}}
\def\s[#1,#2]{[#1\stackrel{\star}{,}#2]}
\def\sx[#1,#2]{[#1\stackrel{\star_{x}}{,}#2]}

\newcommand{\nc}{\newcommand}
\nc{\beq}{\begin{equation}}
\nc{\eeq}{\end{equation}}
\nc{\beqa}{\begin{eqnarray}}
\nc{\eeqa}{\end{eqnarray}}

\def\bc{\begin{center}}
\def\ec{\end{center}}

\def\to{\rightarrow}

\def\gsim{\mathrel{\mathpalette\atversim>}}

\def\bc{\begin{center}}
\def\ec{\end{center}}

\def\gsim{\mathrel{\rlap{\lower4pt\hbox{\hskip1pt$\sim$}}

    \raise1pt\hbox{$>$}}}       

\def\gsim{\mathrel{\rlap{\lower4pt\hbox{\hskip1pt$\sim$}}
    \raise1pt\hbox{$>$}}}       

\begin{document}
\makeatletter
\def\fmslash{\@ifnextchar[{\fmsl@sh}{\fmsl@sh[0mu]}}
\def\fmsl@sh[#1]#2{%
  \mathchoice
    {\@fmsl@sh\displaystyle{#1}{#2}}%
    {\@fmsl@sh\textstyle{#1}{#2}}%
    {\@fmsl@sh\scriptstyle{#1}{#2}}%
    {\@fmsl@sh\scriptscriptstyle{#1}{#2}}}
\def\@fmsl@sh#1#2#3{\m@th\ooalign{$\hfil#1\mkern#2/\hfil$\crcr$#1#3$}}
\makeatother

\thispagestyle{empty}
\begin{titlepage}
\boldmath
\begin{center}
  \Large {\bf  Implications of Quantum Gravity for Dark Matter Searches with Atom Interferometers}
    \end{center}
\unboldmath
\vspace{0.2cm}
\begin{center}
{{\large Xavier~Calmet}\footnote{E-mail: x.calmet@sussex.ac.uk}
{\large and}  {\large  Nathaniel Sherrill}\footnote{E-mail: n.sherrill@sussex.ac.uk}}
 \end{center}
\begin{center}
{\sl Department of Physics and Astronomy,
University of Sussex, Brighton, BN1 9QH, United Kingdom
}\\
\end{center}
\vspace{5cm}
\begin{abstract}
\noindent
In this brief paper, we show that atom interferometer experiments such as MAGIS, AION or AEDGE have the potential to not only probe very light dark matter models, but they will also probe quantum gravity. We show that the linear coupling of a singlet scalar dark matter particle to electrons or photons is already ruled out by our current understanding of quantum gravity coupled to data from torsion pendulum experiments. On the other hand, the quadratic coupling of scalar dark matter to electrons and photons has a large viable parameter space which will be probed by these atom interferometers. Implications for searches of quantum gravity are discussed.
\end{abstract}
\vspace{5cm}
\end{titlepage}



\newpage

A number of experiments based on atom interferometry, see e.g. \cite{Cronin:2009zz}  for a review, are in the planning \cite{Adamson:2018mbw,Abe:2021ksx,Canuel:2017rrp,Canuel:2019abg,Zhan:2019quq,Badurina:2019hst,Badurina:2021rgt,AEDGE:2019nxb}.  MAGIS \cite{Adamson:2018mbw,Abe:2021ksx},  AION \cite{Badurina:2019hst,Badurina:2021rgt} and AEDGE \cite{AEDGE:2019nxb} will provide exciting opportunities to search for gravitational waves and very light new particles.  These experiments rely on a simple physical principle: they are sensitive to fluctuations in the relative phase 
\begin{eqnarray} \label{phase}
\delta\phi= \omega_A \times (2 L),
\end{eqnarray}
between states of cold atom clouds separated by a distance $L$, and $\omega_A$ is the frequency of the atomic transition under consideration. Any interaction of particles with the cold atoms could induce variations  $\delta \omega_A $ in this frequency, 
and the passage of a gravitational wave inducing a strain $h$ would induce a phase shift via 
a change $\delta L = h L$ in the distance of separation. 

Here will focus on the possibility of using atom interferometry experiments to detect extremely light beyond the standard model particles, such as e.g. dark matter, see e.g.\cite{Stadnik:2014tta,Arvanitaki:2016fyj}. These particles can be extremely weakly coupled to standard matter particles. As described in the white papers of these experiments \cite{Adamson:2018mbw,Abe:2021ksx,Badurina:2019hst,AEDGE:2019nxb}, they will be able to cover a large fraction of the, as of now, uncharted parameter space of models of scalar dark matter. Note that MAGIS-100\cite{Adamson:2018mbw,Abe:2021ksx} has very similar sensitivity to that of AION-100 and we shall thus focus our attention on AION-100\cite{Badurina:2019hst}  and AEDGE\cite{AEDGE:2019nxb}.

The generic interactions of a singlet scalar particle $\phi$ with the particles of the standard model that constitute stable matter (electron $\psi_e$, light quarks ($u$, $d$ and $s$-quarks) $\psi_q$, the photon $A_\mu$ and gluons $G^a_\mu$) , can be described by a simple Lagrangian
\begin{eqnarray} \label{lincou}
	\mathcal{L}&=&\kappa  \phi \left (\frac{1}{4}d^{(1)}_e F_{\mu\nu}F^{\mu\nu} -d^{(1)}_{m_e} m_e \bar \psi_e \psi_e \right ) 
	+ \kappa  \phi \left (\frac{1}{4} d^{(1)}_g G_{\mu\nu}G^{\mu\nu} -d^{(1)}_{m_q} m_q \bar \psi_q \psi_q \right )
\end{eqnarray}
with $\kappa=\sqrt{4 \pi G_N}$,  $F_{\mu\nu}=\partial_\mu A_\nu-\partial_\nu A_\mu$ and $G_{\mu\nu}=\partial_\mu G_\nu-\partial_\nu G_\mu-i g_s [G_\mu,G_\nu]$ where $g_s$ is the QCD coupling constant and $G_N$ is Newton constant. We could also add couplings to the field strength of the neutrinos, heavier leptons and quarks, electroweak gauge bosons of the standard model and to the Higgs bosons, but we shall focus here on stable matter which plays the dominant role for very low energy experiments such as atom interferometers.  Note that these operators are dimension 5 operators as they are suppressed by one power of the reduced Planck scale $M_P=1/\sqrt{8 \pi G_N}=2.4\times 10^{18}$ GeV. These linear couplings are the simplest ones possible. The value of the Wilson coefficients $d^{(1)}_j$ depend on the mechanism that generate these operators.  As we use the reduced Planck scale to normalize the operators, $d^{(1)}_j>1$ implies that the interaction generating these operators is stronger that gravitational interactions, while it $d^{(1)}_j<1$, the interactions are weaker than gravitational ones.

Besides the linear coupling in e.q.(\ref{lincou}), a scalar field can couple quadratically to light particles of the standard model via:
\begin{equation}\label{quadcou}
	\mathcal{L}=\kappa^2  \phi^2 \left (\frac{1}{4} d^{(2)}_e F_{\mu\nu}F^{\mu\nu} -d^{(2)}_{m_e} m_e \bar \psi_e \psi_e \right ) 
	+ \kappa^2  \phi^2 \left (\frac{1}{4} d^{(2)}_g G_{\mu\nu}G^{\mu\nu} -d^{(2)}_{m_q} m_q \bar \psi_q \psi_q \right ),
\end{equation}
in other words the interactions of the scalar field with stable matter are suppressed by two powers of the reduced Planck scale. These are non-linear couplings. Higher dimensional couplings are also possible. If the scalar field transforms under some discrete, global or gauge symmetry the linear couplings will in general not be possible  as they violate these symmetries in which case, the simplest coupling to matter is given by the quadratic dimension 6 operators. Note that these operators could account as well for a gauged scalar field, in which case $\phi^2$ would be replaced by $\phi^\dagger \phi$.  Here again, the Wilson coefficients $d^{(2)}_j$ parametrize the strength of the interactions generating these operators.

If the scalar field is sufficiently light, it will approximatively behave as a classical bosonic field
\begin{equation}\label{scalarfield}
\phi(t,\textbf{r}) \simeq A \cos[m_\phi(t - \textbf{v}\cdot \textbf{r}) + \theta],
\end{equation}
where $A$ is the amplitude of the field, $m_\phi$ is its mass and $\theta$ is a phase. This behavior for the scalar field results in sinusoidal time dependence of the electron mass $m_e$, quark masses $m_q$, the fine-structure constant $\alpha$ and the coupling constant of the strong interactions $\alpha_S$ controlled by the couplings $G_N, d_e^{(i)}, d_{m_e}^{(i)},d_q^{(i)}, d_{m_q}^{(i)}$, and the parameters of $\phi(t, \textbf{r})$ in Eq.~\ref{scalarfield}. These oscillations lead to frequency modulations $\delta \omega_A$ and thus phase modulations $\delta \phi$ from Eq.~(\ref{phase}). If we identify this field with dark matter the amplitude is given by 
\begin{equation}
A=\frac{\sqrt{2\rho_{\text{DM}}}}{m_\phi}
\end{equation}
where $\rho_{\text{DM}} \approx 0.4$ GeV/cm$^{3}$ is the local dark matter density. An upper limit on the mass of generic scalar particles, which saturate the local cold dark matter density and form an oscillating classical field, is set by the requirement that there are a large number of such particles within the reduced de Broglie volume. A scalar field lighter than $\sim 0.1$ eV fulfills this constraint \cite{Stadnik:2016vxj}.  For the dark matter wavelength to fit within the smallest known dwarf galaxy, its mass must be larger than $~10^{-22}$ eV.  We shall thus focus on dark matter candidates with masses within the range $10^{-22} \ \mbox{eV} \le m_\phi \le 0.1 \ \mbox{eV}$.

We now come to the main point of this short paper. It has been emphasized \cite{Calmet:2019frv} that whether the Lagrangians in Eqs. (\ref{lincou}) and (\ref{quadcou}) are generated by non-gravitational effects or not, they will be generated by quantum gravitational effects unless forbidden by a symmetry. While the Wilson coefficients generated by perturbative quantum gravity are likely to be small because of loop suppression factors $(16 \pi^2)^{-k} (E/M_P)^{2k}$ where $k$ is the number of loops and $E$ the relevant energy scale for the amplitude, non-perturbative quantum gravity could generate Wilson coefficients of order unity.  While non-perturbative effects in quantum gravity are still poorly understood, there are strong indications that  non-perturbative quantum gravitational effects such as gravitational instantons, wormholes or quantum black holes, see e.g. \cite{Perry:1978fd,Gilbert:1989nq,Chen:2021jcb} will generate an interaction between any scalar field $\phi$ and regular matter with $d_j^{(i)}\sim {\cal O}(1)$ whether such a coupling exists or not when gravity decouples \cite{Calmet:2019jyz,Calmet:2019frv,Calmet:2020rpx,Calmet:2020pub,Calmet:2021iid}. Identical arguments have been made in different contexts, for example in models of grand unification \cite{Ellis:1979fg,Hill:1983xh,Shafi:1983gz,Vayonakis:1993nn}, axion models \cite{Holman:1992va,Holman:1992us,Barr:1992qq,Ghigna:1992iv,Kamionkowski:1992mf}, dark matter models \cite{Calmet:2009uz} or inflationary models \cite{Calmet:2014lga,Kallosh:1995hi}. 

\begin{figure}[H]
	\centering
	\includegraphics[width=8.0cm]{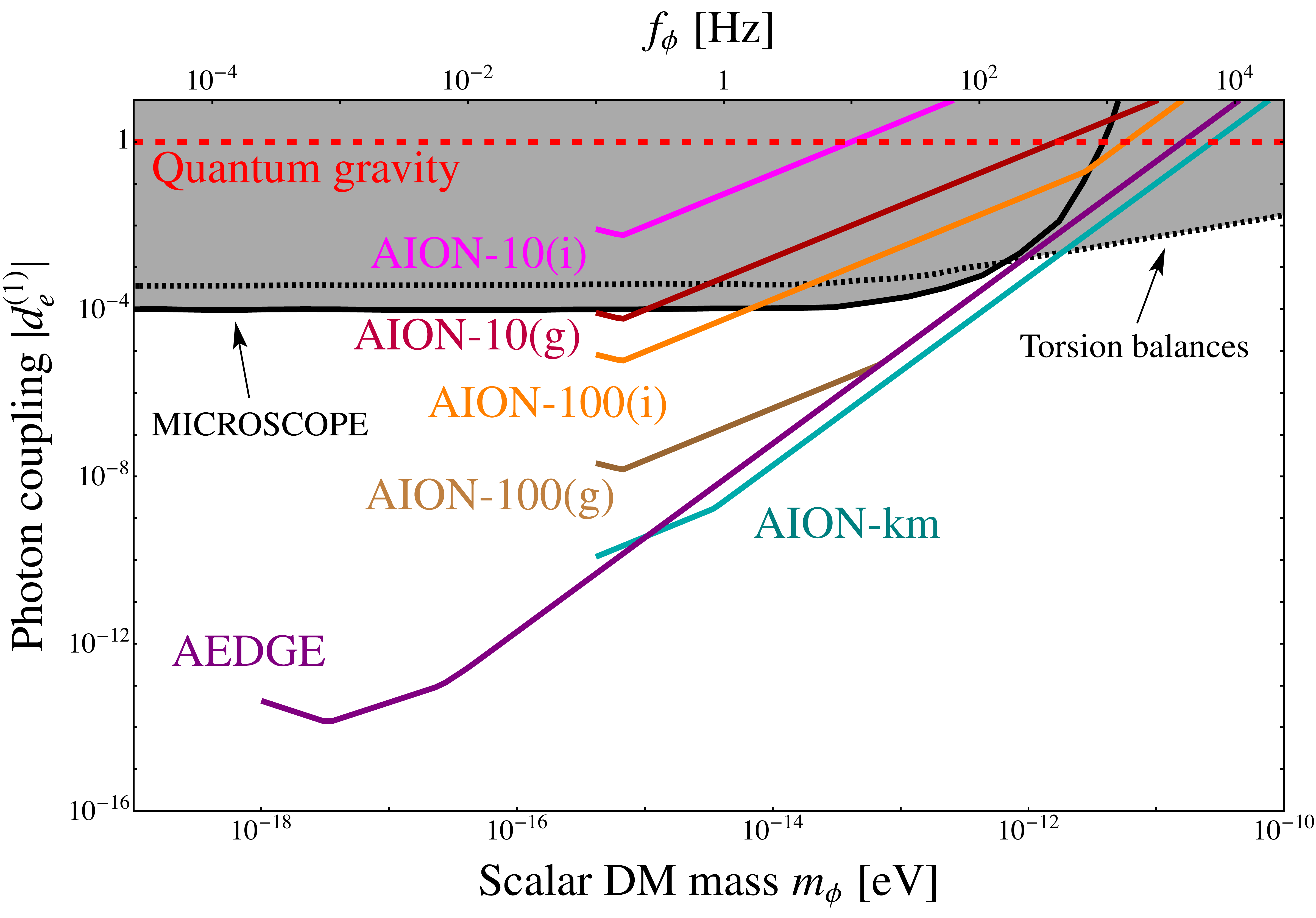}\hspace{2mm} \includegraphics[width=8.0cm]{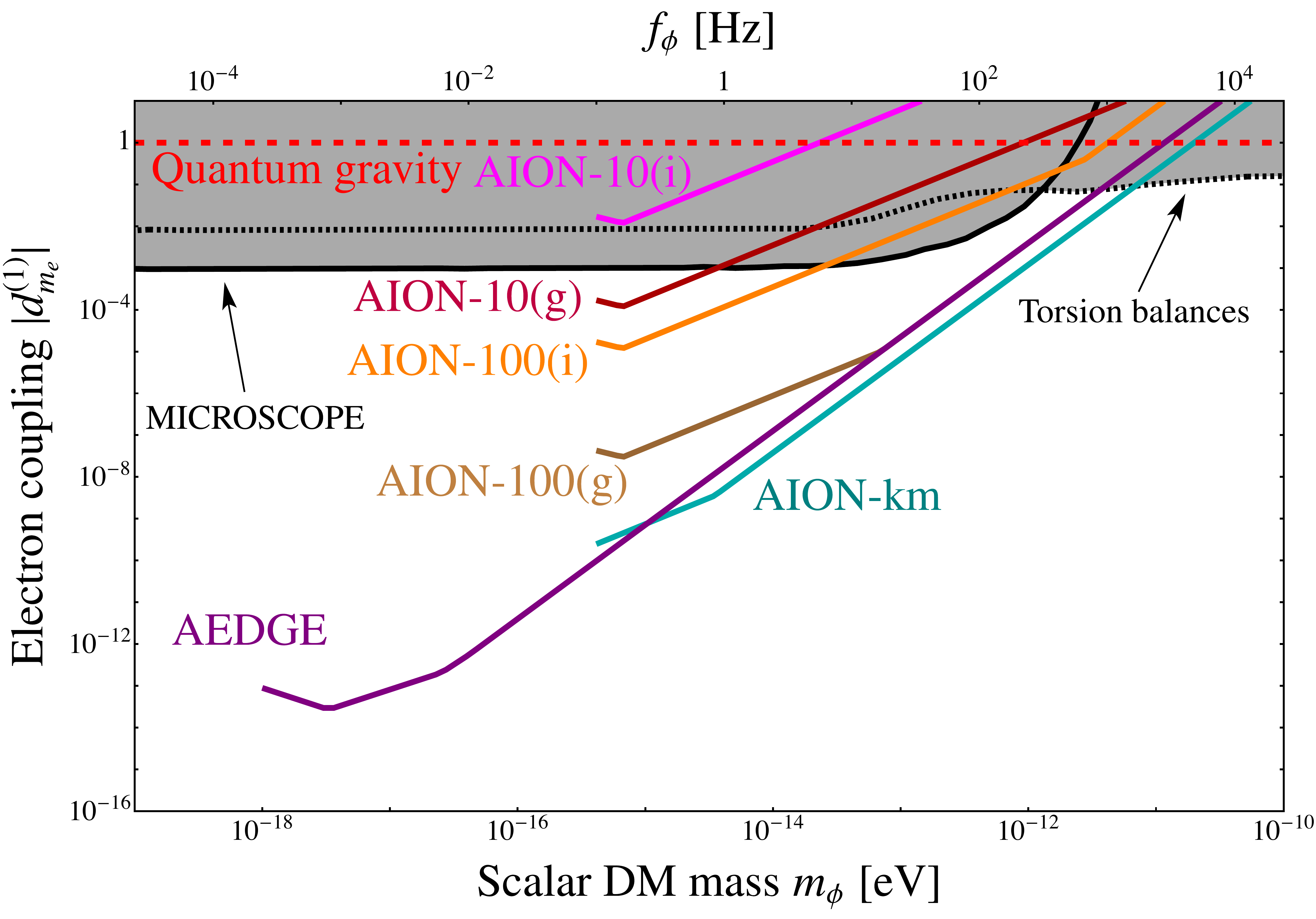}	\caption{Bounds on the parameters $d^{(1)}_e, d^{(1)}_{m_e}$ as a function of the mass of the scalar field. Quantum gravity predicts $d^{(1)}_e, d^{(1)}_{m_e} \sim 1$, closing all of the parameter space accessible to atom interferometers. Note initial (i) and goal (g) sensitivities for AION are displayed.}
	\label{plotde1}
\end{figure}

However, very light scalar fields coupling linearly to regular matter (i.e. dimension five operators) are essentially ruled out by the E\"ot-Wash torsion pendulum experiment \cite{Kapner:2006si,Hoyle:2004cw,Adelberger:2006dh,Lee:2020zjt}  for $d_j^{(1)}\sim {\cal O}(1)$. Indeed, E\"ot-Wash's data implies that if  $d_j^{(1)}\sim1$, the mass of the singlet scalar field must be heavier than $10^{-2}$ eV \cite{Calmet:2019jyz,Calmet:2019frv,Calmet:2020rpx,Calmet:2020pub,Calmet:2021iid}. We see from  
Figure (\ref{plotde1}) that all of the parameter space that can be probed with atom interferometers is ruled out if indeed quantum gravity produces these operators and given the limit from E\"ot-Wash.  Atom interferometers will therefore provide a very important test of quantum gravity. If a very light neutral scalar field with linear coupling to regular matter was found by such experiments with a coupling constant much smaller than one, we would learn that dimension 5 operators are not generated by quantum gravity. 

Note that we have not yet considered the coupling $d^{(1)}_H \phi H^\dagger H$ which has been investigated in \cite{Adamson:2018mbw,Badurina:2019hst,AEDGE:2019nxb}.  If this operator is generated by quantum gravity, it must vanish in the limit $M_P \to \infty$. There are thus two types of operators that could be generated:  $d^{(1)}_H=m \exp{(-M_P/\mu)}$ where $\mu$ is some renormalization scale of the order of the atomic scale and $m$ could be the mass of the dark matter scalar field, the mass of the Higgs field or the Planck mass or $d^{(1)}_H=m_1 (m_2/M_P)^n$ where $m_{1/2}$ can be the mass of the dark matter scalar field or that of the Higgs boson. In the exponential suppression case, the operator is clearly very suppressed and irrelevant. In the power of Planck mass suppression case, the operator is also very suppressed if one of the $m_i$ is the mass of the dark matter particle. However, it could be observable if it was the Higgs boson mass for $n=1$, in that case quantum gravity predicts $d^{(1)}_H\sim 7 \times 10^{-6}$ eV which is excluded by torsion balance experiments for all dark matter masses relevant to atom interferometers as can be seen from Figure ({\ref{plotdH}).
\begin{figure}[H]
	\centering
	\includegraphics[width=12cm]{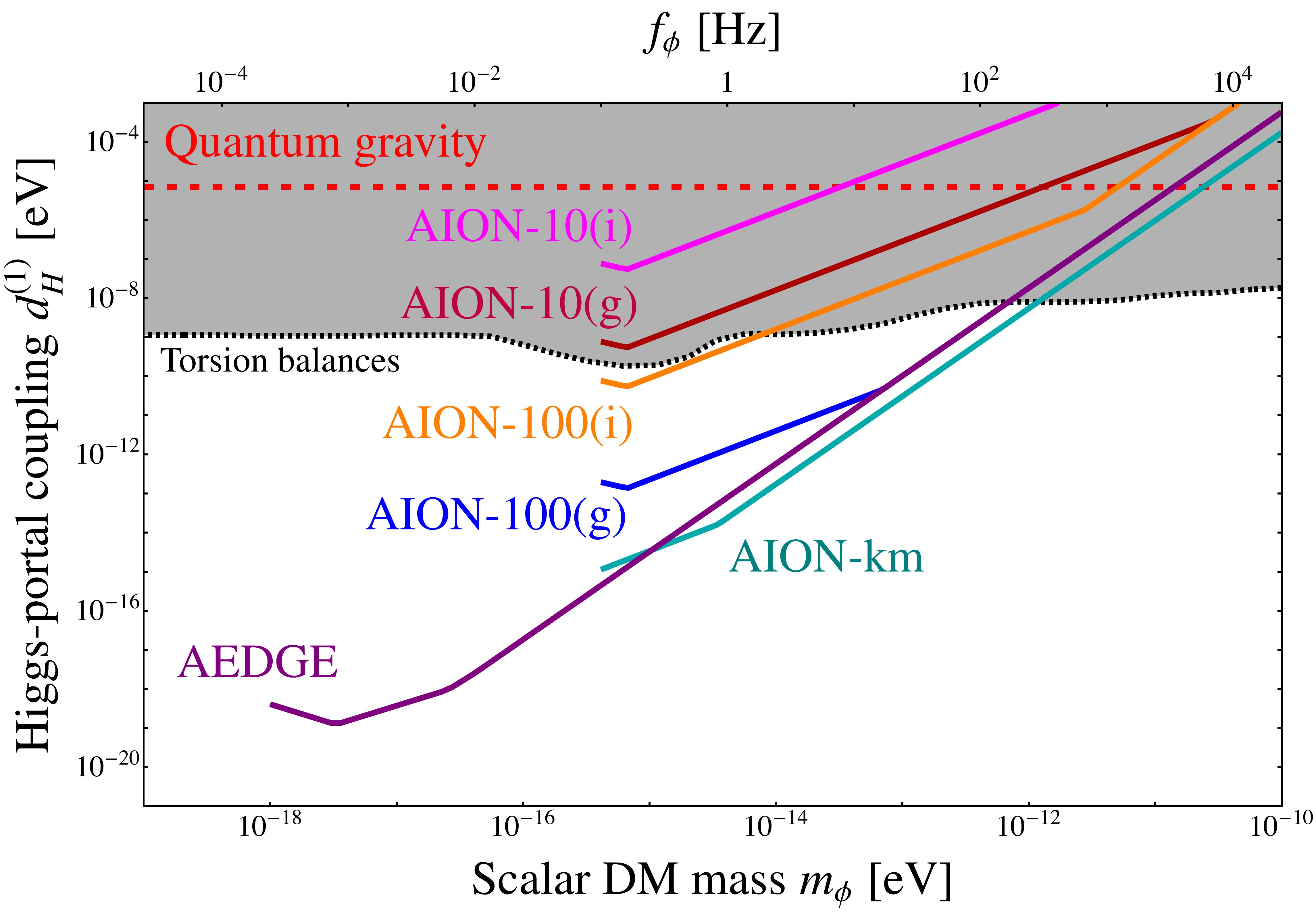}
	\caption{Bounds on the parameters $d^{(1)}_H$. Quantum gravity predicts $d^{(1)}_H\sim 7 \times 10^{-6}$ eV.}
	\label{plotdH}
\end{figure}
Furthermore, note that atom interferometers are not sensitive to the couplings $d^{(i)}_g$ and $d^{(i)}_{m_q}$, i.e. the couplings of scalar fields to quarks and gluons, but only to the couplings of scalar fields to electrons or to the photon. While quantum gravity predicts $d^{(i)}_g \sim 1$ and $d^{(i)}_{m_q} \sim 1$, these predictions are not relevant for these experiments.

While linear couplings are essentially ruled out, on the other hand, as can be seen from Figure (\ref{plotde2}),  the quadratic couplings and other non-linear couplings are far less constrained by quantum gravity and atom interferometers  will be able to explore uncharted territory and test non-gravitational interactions between very light scalar fields and dark matter. Looking at Figure (\ref{plotde2}),  we see that the quadratic coupling induced by quantum gravity of a scalar field dark matter particle to photons or electrons could be discovered by AEDGE for a scalar mass heavier than $\sim 5 \times 10^{-17}$ eV and lighter than $10^{-21}$ eV. This lower bound comes from the requirement that the dark matter scalar field wave function fits into the smallest known galaxies. In that sense AEDGE has the potential of testing quantum gravity if such very light scalar fields exist in nature. It could discover not only dark matter but also quantum gravity at the same time.

\begin{figure}[H]
	\centering
	\includegraphics[width=8cm]{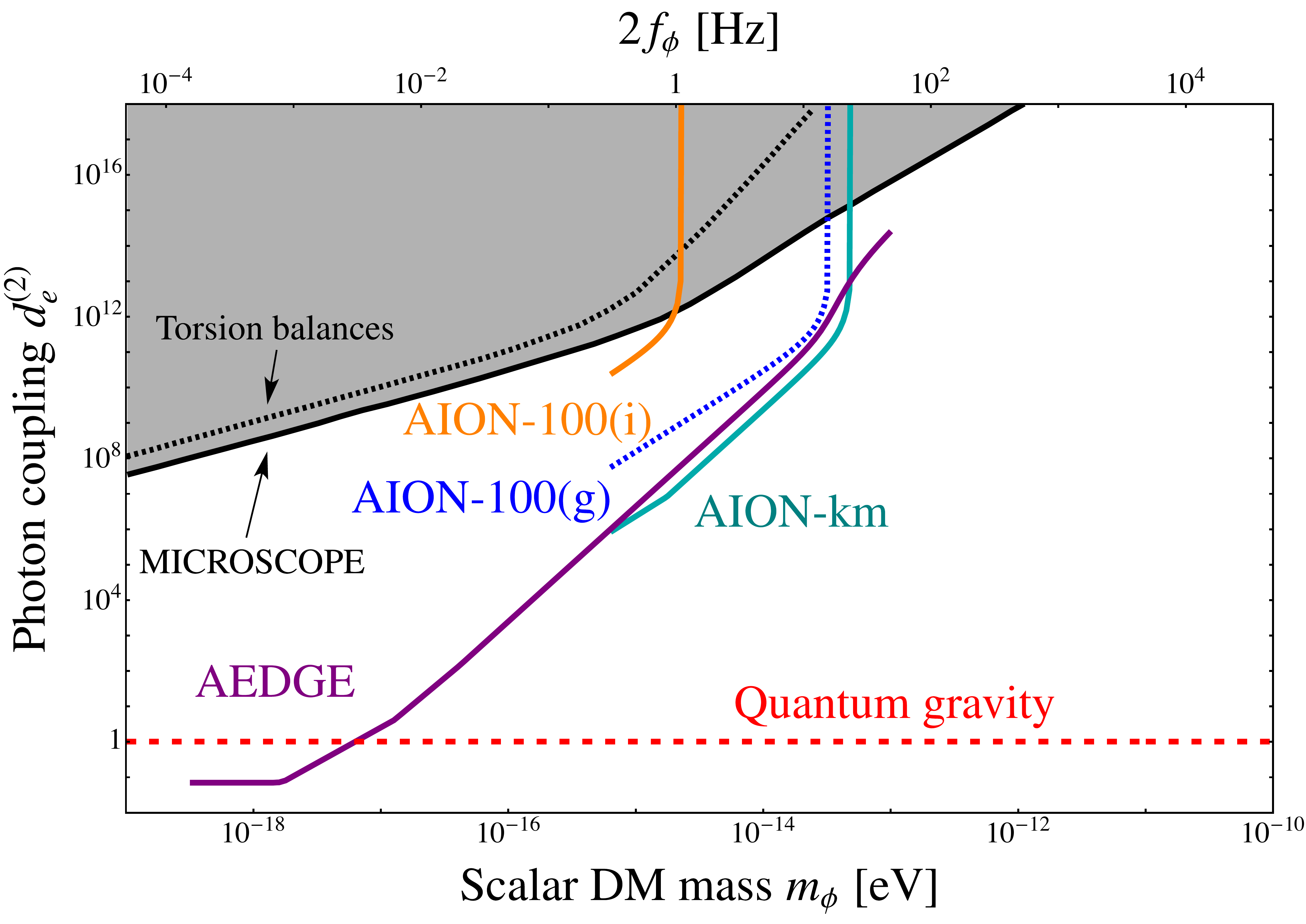}\hspace{2mm}  \includegraphics[width=8cm]{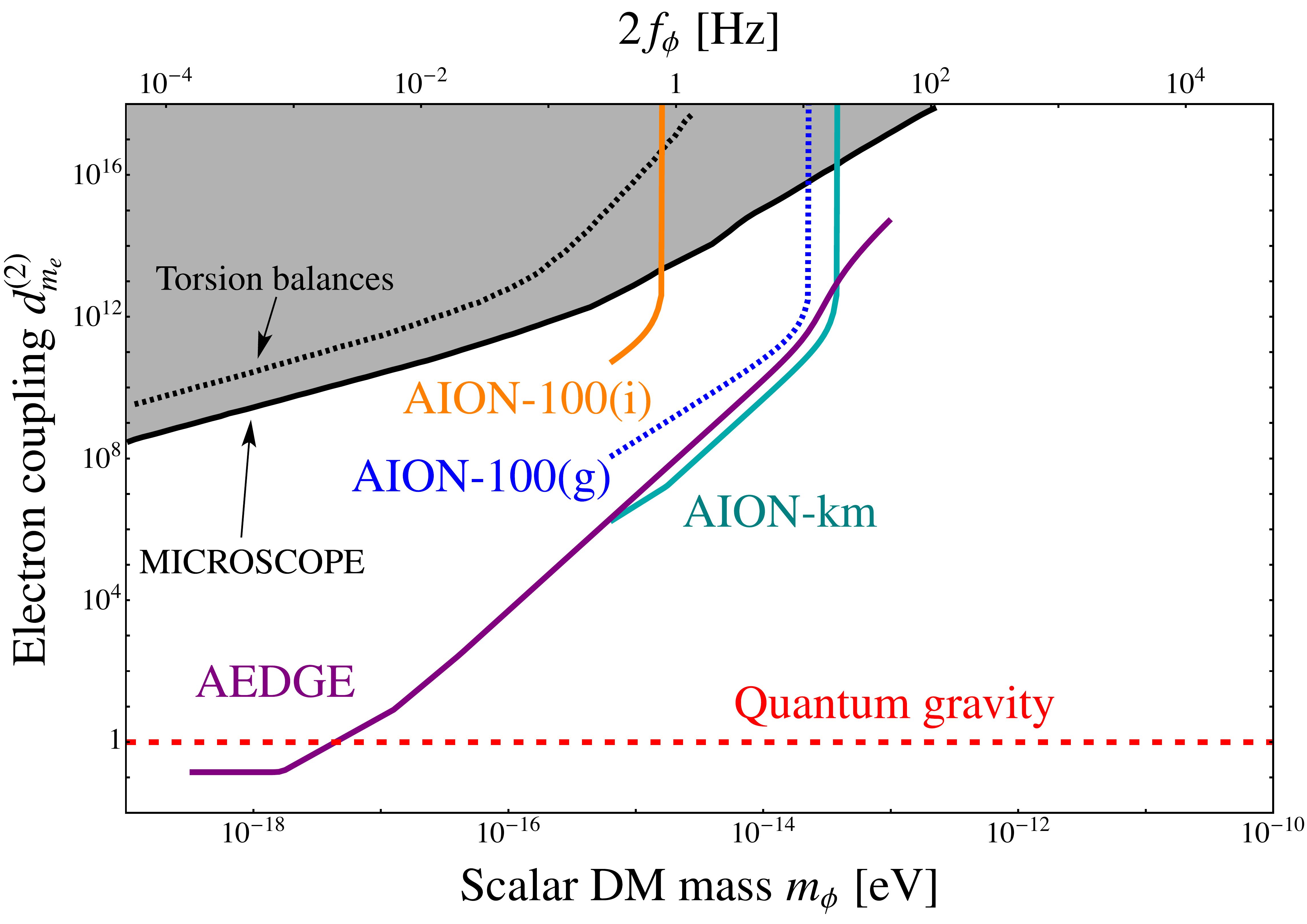}
	\caption{Bounds on the parameters $d^{(2)}_e, d^{(2)}_{m_e}$ as a function of the mass of the scalar field. Quantum gravity predicts $d^{(2)}_e,  d^{(2)}_{m_e} \sim 1$. AEDGE could detect a quantum-gravitational interaction of scalar dark matter with electrons for dark-matter masses below $\sim 5 \times 10^{-17}$ eV.}
	\label{plotde2}
\end{figure}

Finally let us emphasize that we have considered here a very conservative model of quantum gravity with a scale of quantum gravity around $2.4\times 10^{18}$ GeV which is the traditional reduced Planck scale. However, in models with large extra-dimensions \cite{Arkani-Hamed:1998jmv,Antoniadis:1998ig,Randall:1999ee} or a large number of particle in a hidden sector \cite{Calmet:2008tn}, the scale of quantum gravity could be well within the TeV region.  Searches at the Large Hadron Collider at CERN,  see e.g. \cite{CMS:2018ozv}, indicate that there are no quantum black holes with masses below 10 TeV depending on the specific quantum black hole model, so we can conservatively assume a lower bound for the scale of quantum gravity of the order of 10 TeV. In these models, the Wilson coefficients of the effective action discussed in Eqs.  (\ref{lincou}) and  (\ref{quadcou}) could be as large as  $d^{(1)}_{e/m_e}\sim 2\times 10^{14}$ for the linear couplings which accounts for an effective Planck mass of $10^4$ GeV
and $d^{(2)}_{e/m_e}\sim 4 \times 10^{28}$ for the quadratic couplings which again accounts for an effective Planck mass of $10^4$ GeV. We thus see that atom interferometers such as AION and MAGIS will probe a variety of quantum gravitational models already with the initial configurations.  To put things in perspective, a $d^{(2)}_{e/m_e}\sim 1\times 10^{8}$ in Figure (\ref{plotde2}) corresponds to a scale of quantum gravity of $2 \times 10^{10}$ GeV. AION and AEDGE will thus probe a wide range of models of quantum gravity. 

In this short paper, we have shown that quantum gravity can be probed with atom interferometers built to detect gravitational waves or very light dark matter particles. We have shown that quantum gravity essentially rules out a linear coupling of dark matter scalar fields to electrons or photons as it would have already been seen by torsion pendulum experiments such as E\"ot-Wash or MICROSCOPE \cite{MICROSCOPE:2019jix} in space. On the other hand, non-gravitational quadratic couplings of scalar dark matter could be discovered by future interferometers such as MAGIS or AION. AEDGE even has the potential to discover simultaneously dark matter and quantum gravity.

{\it Acknowledgments:}
    We would like to thank Oliver Buchmuller,  Ian Shipsey and Christopher McCabe in particular for providing us with the data used to make their plots in \cite{Badurina:2019hst}.
This work is supported in part  by the Science and Technology Facilities Council (grants numbers ST/T00102X/1, ST/T006048/1 and ST/S002227/1). 

\end{document}